\documentclass[12pt,preprint]{aastex}
\usepackage{graphicx}
\usepackage{natbib}

\shorttitle{Gravitational Lensing by Black Holes}

\shortauthors{V. Bozza and L. Mancini}

\begin{document}
\title{Gravitational Lensing by Black Holes: a
comprehensive treatment and the case of the star S2}

\author{V. Bozza$^{1,2,3}$ and L. Mancini$^{2,3,4}$}

\affil{$^1$Centro Studi e Ricerche "Enrico Fermi", via Panisperna
89/A, Roma, Italy.}

\affil{$^2$Dipartimento di Fisica "E.R. Caianiello", Universit\'a
di Salerno, via S. Allende, Baronissi (SA), Italy.}

\affil{$^3$Istituto Nazionale di Fisica Nucleare, Sezione di
Napoli, Italy.}

\affil{$^4$Institut f\"{u}r Theoretische Physik der
           Universit\"{a}t Z\"{u}rich, CH-8057 Z\"{u}rich, Switzerland}

\date{Received  / Accepted}

\begin{abstract}

Light rays passing very close to a black hole may experience very
strong deviations. Two geometries were separately considered in
the recent literature: a source behind the black hole (standard
gravitational lensing); a source in front of the black hole
(retro-lensing). In this paper we start from the Strong Field
Limit approach to recover both situations under the same
formalism, describing not only the two geometries just mentioned
but also any intermediate possible configurations of the system
source-lens-observer, without any small-angle limitations. This is
done for any spherically symmetric black holes and for the
equatorial plane of Kerr black holes. In the light of this
formalism we revisit the previous literature on retro-lensing,
sensibly improving the observational estimates. In particular, for
the case of the star S2, we give sharp predictions for the
magnitude of the relativistic images and the time of their highest
brightness, which should occur at the beginning of year 2018. The
observation of such images would open fascinating perspectives on
the measure of the physical parameters of the central black hole,
including mass, spin and distance.

\end{abstract}

\section{Introduction}

The fact that light rays can wind an arbitrary number of times
around a black hole before emerging back towards spatial infinity
is well known since early times of General Relativity
\citep{Dar,Atk,Lum,Oha,Nem} (see e.g. the treatment in
\citet{Cha}). In practice, a source behind a black hole not only
produces the two classical weak field gravitational lensing
images, but also an infinite number of strong field images
corresponding to photons with winding numbers running from 1 to
infinity. This phenomenon has been revived in a work by
\citet{VirEll}, who showed that the supermassive black hole at the
center of our Galaxy may be a suitable lens candidate. The
resolution needed for such observation is very high but should be
reached by next future Very Long Baseline Interferometry
experiments, such as ARISE\footnote{ARISE web page:
arise.jpl.nasa.gov} and MAXIM\footnote{MAXIM web page:
maxim.gsfc.nasa.gov}. In order to describe gravitational lensing
in such extreme cases, we cannot use any weak field approximation.
However, it is possible to take advantage of the opposite limit
and finally get a very simple and efficient analytical
approximation, called the Strong Field Limit (SFL) \citep{Boz1}.
This method, firstly emerged in Schwarzschild black hole studies
\citep{Dar,Oha,BCIS}, was then applied to Reissner-Nordstrom black
holes \citep{ERT}, to charged black holes of heterotic string
theory \citep{Bha} and finally generalized to an arbitrary
spherically symmetric metric \citep{Boz1}. The microlensing
situation was considered by \citet{Pet}, while the Kerr black hole
was explored analytically for quasi-equatorial motion \citep{Boz2}
and numerically for arbitrary motion \citep{VazEst}. It is also
interesting that a time delay measurement would give a precise
estimate of the distance of the lens \citep{BozMan}. For former
and/or alternative formulations of strong field gravitational
lensing see \citet{FriNew,FKN,DabSch,Per}.

Already in the first studies of gravitational lensing in strong
fields by \citet{Dar}, it was noticed that the whole sky is mapped
in the vicinity of what it was then called a "compact sun". In
particular, it was known that a source in front of the lens could
yield relativistic images and Einstein rings as bright as a source
behind the lens (see e.g. \citet{Lum}). In more recent years
retro-lensing was re-discovered quite independently from standard
gravitational lensing. \citet{HolWhe} proposed that a black hole
of a few solar masses passing within 1 pc from the solar system
would redirect photons from the Sun backwards to the Earth. An
observer would thus see a "star" lighting up and then switching
off in the sky as the Earth in its motion enters and leaves the
best alignment position. \citet{DepS2} suggested the black hole in
Sgr A* at the center of our Galaxy as a suitable retro-lens,
proposing the star S2 (the star with the smallest average distance
from Sgr A* discovered so far) as a candidate source. At the
moment, this seems to be the best known candidate for
gravitational lensing in strong fields, deserving a closest
investigation. At the same time, \citet{EirTor} considered
retro-lensing by Sgr A*, using the analytical framework of the SFL
method to give correct estimates for such a phenomenon; their
investigation, however, is limited to small angles and cannot
cover the case of S2. Finally, \citet{DepRM} extended Holz \&
Wheeler proposal to Kerr retro-lensing, using the SFL method to
calculate the position of the images but using the Holz \& Wheeler
formula for the amplification (which is inadequate for spinning
black holes), finding no significant deviation from the light
curves of the Schwarzschild case.

In this paper, we give up any limitation due to small-angle
approximation and treat the standard and retro-gravitational
lensing on a unified ground, where they just come up as particular
cases. Besides recovering these two situations, we also address
the gravitational lensing problem for any intermediate geometries,
where the strong field images are still present as always, but
they are usually dimmer (see Sect. 2). This allows us to treat in
a very accurate way the case of the star S2, giving sharp
predictions for the light curves of its images in the next years.
Moreover, we clarify several aspects which were not clearly stated
in the former literature, constructing a unique analytical
framework for the whole phenomenology.

In Sect. 2 we treat the spherically symmetric lens; we give the
formulae for the position and the magnification of the images, and
discuss the differences with the magnification formula by
\citet{HolWhe}; we also comment on the importance of time delay
measurements. Sect. 3 is devoted to the study of S2, the best
candidate source for gravitational lensing in the SFL; we use our
formalism to draw analytical curves of the images, accurately
predicting the epoch of their luminosity peak. In Sect. 4, in the
light of the results by \citet{Boz2}, we guess about the possible
changes in S2 relativistic images if the black hole at the center
of our galaxy has non-vanishing spin; we also compare our results
in Sun retro-lensing by Kerr black holes with those by
\citet{DepRM}. Sect. 5 contains a summary of the work.

\section{Gravitational lensing by spherically symmetric black holes}

According to the SFL method, the deflection angle of a light ray
passing very close to a black hole can be expanded around the
minimum impact angle $\theta_m$, which separates the light rays
absorbed by the black hole ($\theta<\theta_m$) from the light rays
which are simply deflected ($\theta>\theta_m$). As previously
shown \citep{Boz1}, the deflection angle always diverges
logarithmically at $\theta \sim \theta_m$ for any class of
spherically symmetric black holes. The logarithmic term and the
constant term give a sufficient approximation to the deflection
angle in order to explain the whole phenomenology. The fundamental
formula reads

\begin{equation}
\alpha(\theta)=-\overline{a} \log \left(\frac{\theta}{\theta_m}-1
\right) +\overline{b}, \label{SFL}
\end{equation}
up to higher order terms in $(\theta-\theta_m)$. The numerical
coefficients $\overline{a}$ and $\overline{b}$ depend on the
characteristics of the black hole (electric charge, coupling to a
scalar field, the particular gravitation theory we are using,
etc.). We refer the reader to \citep{Boz1} for its full derivation
and for some examples (see also \citet{Bha}).

Using the formula (\ref{SFL}), \citet{EirTor} have correctly
calculated the position and the magnification of the retro-lensing
images. However, their treatment (as well as the treatment in
\citep{Boz1} for standard lensing) is limited to small separations
of the source from the optical axis (defined as the line
connecting the observer with the lens). In order to address the
general case, we have to write the lens equation in a suitable
way, without restricting to particular cases.

\begin{figure}
\resizebox{\hsize}{!}{\includegraphics{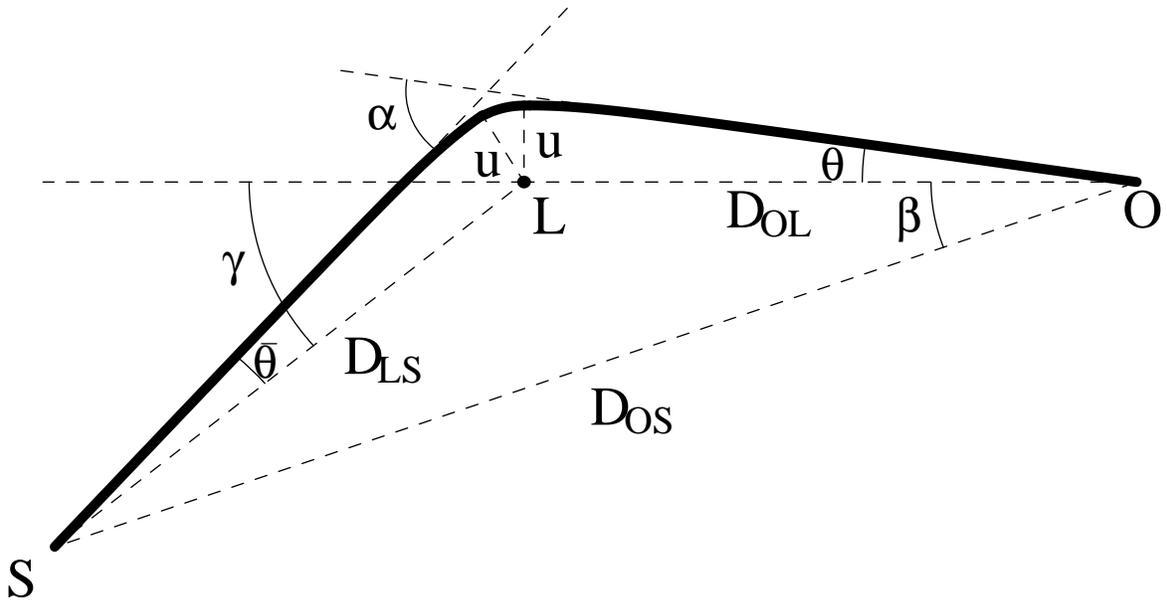}}
 \caption{A generic lensing configuration.}
 \label{Fig Lens Eq}
\end{figure}

\subsection{Lens equation and position of the images}

The source, the lens and the observer define the plane where the
whole motion of the photon takes place in the case of spherically
symmetric black holes. We define $\gamma$ as the angle between the
source-lens direction and the optical axis (see Fig. \ref{Fig Lens
Eq}). $\theta$ is the angular position of the image in the sky of
the observer, with respect to the position of the black hole. It
coincides with the impact angle of the light ray as seen by the
observer. The impact angle as seen from the source is called
$\overline{\theta}$. Then the lens equation is simply (see also
\citet{Boz2})
\begin{equation}
\gamma=\alpha(\theta)- \theta -\overline{\theta}. \label{Lens Eq}
\end{equation}
Now, we can give a unique treatment allowing $\gamma$ to run over
the whole range $[0,+\infty)$. In fact, $\gamma \simeq 0$ would
yield the weak field gravitational lensing; $\gamma \simeq \pi$
would be gravitational retro-lensing; $\gamma \simeq 2\pi$ would
give standard gravitational lensing in the SFL and so on. The
collection of strong field gravitational lensing images would be
recovered solving the lens equation for $\gamma \simeq 2n \pi$
with $n \geq 1$, while the collection of retro-lensing images is
recovered with $\gamma=(2n-1)\pi$ with $n\geq 1$.

Since we are only interested in strong field gravitational lensing
of far sources, the impact angle $\theta=\arcsin (u/D_{OL})$ is
always negligible compared to $\gamma$ and $\alpha$, since it is
of the order of the Schwarzschild radius of the black hole divided
by the distance of the lens which is usually much larger (unless
we are falling into the black hole!). So we can safely drop
$\theta$ from Eq. (\ref{Lens Eq}). For $\overline{\theta}$ we can
apply the same argument as long as the source is far enough from
the black hole (see e.g. \citet{CunBar} or \citet{Vie} for the
more complicated case where the source is orbiting very close to
the black hole).

The general solution of the lens equation is then
\begin{equation}
\theta=\theta_m \left(1+ e^{\frac{\overline{b}-
\gamma}{\overline{a}}} \right). \label{theta}
\end{equation}
This formula is valid both for standard gravitational lensing
(when $\gamma \simeq 2n\pi$, compare with \citet{Boz1}) and for
retro-lensing (when $\gamma \simeq (2n-1) \pi $, compare with
\citet{EirTor}). It is also valid for any intermediate situation.
Its limits of validity are fixed by the accuracy of the SFL
formula (\ref{SFL}) for the deflection angle when we slip from
strong field to weak field gravitational lensing. This point
deserves a deeper discussion.

\begin{figure}
\resizebox{\hsize}{!}{\includegraphics{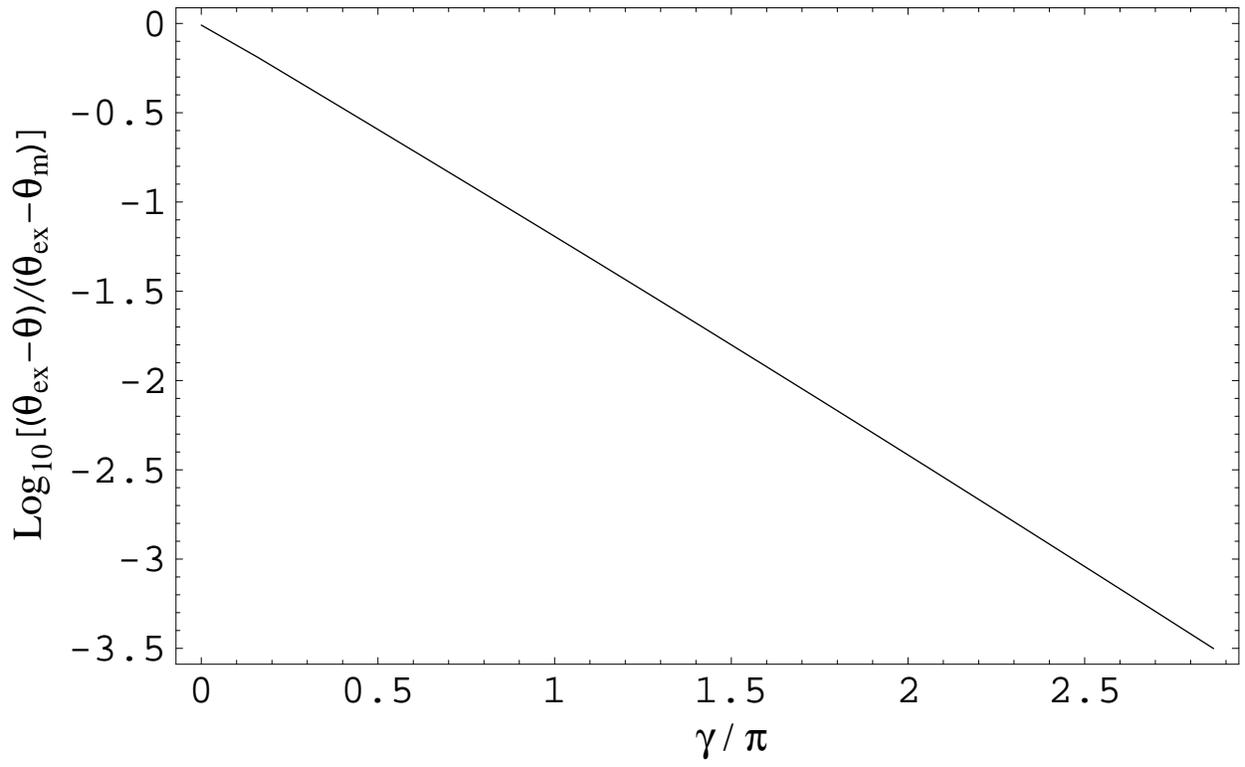}}
 \caption{Accuracy of the SFL in the determination of the images as a function of the source position angle $\gamma$.}
 \label{Fig acc}
\end{figure}

Estimates of the accuracy of the SFL approximation are easily done
in the Schwarzschild case, where it is possible to calculate the
deflection angle exactly \citep{BCIS}. So let us call
$\alpha_{ex}(\theta)$ the exact deflection angle for Schwarzschild
black hole. Let $\theta_{ex}$ be the image calculated using
$\alpha_{ex}$ in the lens equation (\ref{Lens Eq}). What is really
interesting in the images is their difference from the minimum
impact angle, i.e. $\theta-\theta_m$ and $\theta_{ex}-\theta_m$.
So, in Fig. \ref{Fig acc}, we finally plot
$(\theta_{ex}-\theta)/(\theta_{ex}-\theta_m)$ vs $\gamma$.

Of course when $\gamma \simeq 0$ we should rather use the weak
field approximation and the error becomes very large (we should
also take into account the fact that the additional terms in the
lens equation are no longer negligible). What is more interesting
is looking at the error in the determination of the retro-lensing
images ($\gamma \simeq \pi$), which turns to be of the order of
$6\%$. The accuracy grows exponentially with $\gamma$ so that the
outermost relativistic image of standard strong field
gravitational lensing ($\gamma=2\pi$) is determined with an
accuracy of $0.3\%$. To improve these numbers, which is perhaps
desirable for the outermost retro-lensing image, higher order
terms in the formula for the deflection angle should be included,
but this task is beyond the scope of this work. A $6\%$ accuracy
in retro-lensing will be sufficient for most of the following
discussions.

\subsection{Magnification of the images}

In the first retro-lensing paper \citep{HolWhe} a rather unusual
formula for the amplification of the images was proposed, based on
heuristic arguments
\begin{eqnarray}
&& A=\pi (u^2_o-u^2_i) \frac{\Delta \Theta}{2\pi} \\
&& \Delta \Theta= 2 \arctan \left( \frac{R_S}{D_{OS} \sin \beta}
\right),
\end{eqnarray}
where $u_i$ and $u_o$ are the inner and outer impact parameters
which bound the image, $\Delta \Theta$ is the amplitude of the arc
described by the image, $R_S$ is the source radius.

According to this formula, a full Einstein ring would form only
for a perfect alignment $\beta=0$. This is obviously wrong, since
it is sufficient that just one point of the source is perfectly
aligned to build a full Einstein ring. For a source far enough
from perfect alignment, it can be shown that this approximate
formula coincides with the standard formula we are going to derive
in this subsection. However, it cannot be trusted in the high
alignment regime and should be replaced by a more rigorous one.

\citet{Oha} already derived a formula valid for the Schwarzschild
black hole. Here we rederive it for an arbitrary black hole. To
describe the extension of the source in terms of our angles, we
can attach polar coordinates to the lens frame, where $\gamma$
plays the role of the polar angle from the optical axis and we
introduce an azimuthal angle $\phi$ around this axis. Then the
solid angle element is given by $d\omega_S=d\gamma \sin \gamma
d\phi$. In the same way, we can attach polar coordinates to the
observer obtaining an image element $d\omega_I=d\theta \sin \theta
d\phi$. The quantity
\begin{equation}
\frac{d\omega_I}{d\omega_S}=\frac{\sin\theta}{\sin{\gamma} \frac{d
\gamma}{d\theta}}
\end{equation}
represents the ratio between the angular extension of the image as
it appears to the observer and the extension of the source as it
appears to the black hole. The latter is related to the extension
of the source as it appears to the observer by the simple
quadratic ratio of the distances $D_{OS}^2/D_{LS}^2$. Finally we
have
\begin{equation}
\mu=\frac{D_{OS}^2}{D_{LS}^2}\frac{\sin\theta}{\sin{\gamma}
\frac{d \gamma}{d\theta}}. \label{mu}
\end{equation}

The source distance is of course related to the other distances by
simple trigonometry: $D_{OS}^2=D_{OL}^2+D_{LS}^2+2D_{OL}D_{LS}
\cos \gamma$. Moreover, since we are always considering black
holes far from the observer, we can safely approximate $\sin\theta
\simeq\theta$. On the other hand, only for a high alignment
situation, with $\gamma\simeq k\pi$, we can replace $\sin \gamma$
by $|\gamma-k\pi|$. In such cases, which correspond to standard
and retro-lensing, we recover the usual formula for the
magnification of spherically symmetric lenses, valid for small
angles. Eq. (\ref{mu}), on the contrary, is valid also for a
general $\gamma$.

Now, using the lens equation, we can write
\begin{equation}
\mu=-\frac{D_{OS}^2}{D_{LS}^2} \frac{\theta_m^2
e^{\frac{\overline{b}- \gamma}{\overline{a}}}(1+
e^{\frac{\overline{b}- \gamma}{\overline{a}}})}{\overline{a} \sin
\gamma},
\end{equation}
where the sign coming out of the derivative correctly accounts for
the parity of the image. For $\gamma \simeq 2n\pi$ we recover the
magnification formula given by \citet{Boz1}, while for $\gamma
\simeq (2n-1)\pi$ we recover the formula given by \citet{EirTor}.
However, this formula smoothly joins the two extreme cases,
covering the whole range of $\gamma$. It can be obtained from the
Kerr magnification formula given by \citet{Boz2} in the limit of
vanishing black hole spin. However, the derivation proposed here
does not need to pass through the Kerr metric and can be applied
to a generic spherically symmetric black hole.

The caustic points (points of formally infinite magnification) are
exactly at $\gamma=k\pi$ where $\sin \gamma$ vanishes. Sources
close to these points get the maximal amplification. That is why
standard and retro-lensing were first studied as the most
interesting physical cases.

To treat extended sources we need to integrate this formula over
the angular extension of the source, eventually weighted by a
surface brightness factor. If the source is far from a caustic
point, then the magnification varies very little along the source
surface and it makes almost no difference to approximate the
source as a point. On the contrary, the magnification for
point-like sources diverges on caustic points, while for realistic
extended sources we get a finite result integrating over the
source angular area. Then the source radius acts as an effective
cut-off of the magnification.

In order to compare this formula with that of \citet{HolWhe}, we
can use it in the limit of retro-lensing, with $\gamma \simeq
\pi$. The magnification becomes
\begin{equation}
\mu_{retro}=-\frac{D_{OS}^2}{D_{LS}^2} \frac{\theta_m^2
e^{\frac{\overline{b}- \pi}{\overline{a}}}(1+
e^{\frac{\overline{b}- \pi}{\overline{a}}})}{\overline{a}
|\gamma-\pi|},\label{mur}
\end{equation}
where we have neglected the dependence on $\gamma$ in the
exponentials being much weaker than the dependence in the
denominator.

Finally, the integral over the source extension, for a constant
brightness source of radius $R_S$ is a standard calculation in
terms of elliptic functions, first performed by \citet{WitMao},
and also given explicitly by \citet{EirTor}. It amounts to
replacing the $|\gamma-\pi|^{-1}$ in Eq. (\ref{mur}) by its
integral over the source disk $D_S$
\begin{eqnarray}
&\frac{1}{\pi \gamma_S^2}&\int\limits_{D_S} \frac{1}{\gamma}\sin
\gamma
d\gamma d\phi= \frac{2Sign[\gamma_S-\gamma]}{\pi \gamma_S^2} \nonumber \\
&& \left[ (\gamma_S-\gamma)  E\left(\frac{\pi}{2},-\frac{4\gamma_S
\gamma}{(\gamma_S-\gamma)^2} \right)+\right. \nonumber \\
 && \left.
(\gamma_S+\gamma)F\left(\frac{\pi}{2},-\frac{4\gamma_S
\gamma}{(\gamma_S-\gamma)^2} \right) \right],\label{Intmu}
\end{eqnarray}
where for simplicity we have used $\gamma$ instead of
$|\gamma-\pi|$ and we have defined $\gamma_S=R_S/D_{LS}$. $F$ and
$E$ are the elliptic integrals of first and second kind.

\begin{figure}
\resizebox{\hsize}{!}{\includegraphics{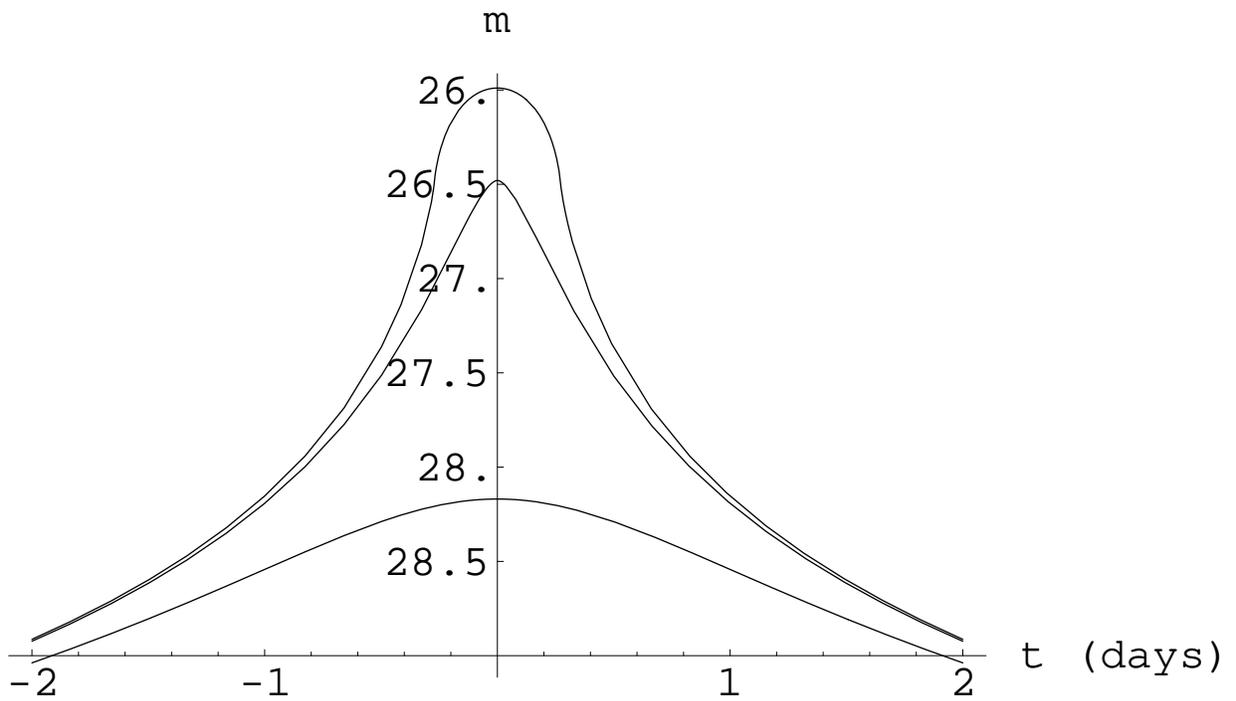}}
 \caption{Total magnitude of the images of the Sun produced by a black hole with 10 M$_\odot$ at 0.01 pc distance
 acting as a retro-lens. From bottom to top we consider a central event, an event with maximum alignment of
 $\gamma=1R_\odot/D_{LS}$ and $\gamma=1 AU/D_{LS}$.}
 \label{Fig Micro}
\end{figure}

With this formula, we can draw a retro-MACHO light curve for the
situation described by \citet{HolWhe}, i.e. a black hole of 10
M$_\odot$ passing at 0.01 pc from the Sun, which deviates photons
coming from the Sun backwards to the Earth. We can compare the
curves in Fig. \ref{Fig Micro} with those by \citet{HolWhe} to see
how different the shape looks when the source starts to cover the
caustic point (top curve). It is interesting to note that the
absolute normalization of the curves by \citet{HolWhe} is slightly
different from that by \citet{DepRM} and both are slightly
different from ours. This might depend on minor approximations in
the units of measure.

In the case of perfect alignment, Eq. (\ref{Intmu}) gives
$2/\gamma_S$. The maximum amplification is thus
\begin{equation}
\mu_{max}=\frac{D_{OS}^2}{D_{LS}^2} \frac{4\theta_m^2
e^{\frac{\overline{b}- k\pi}{\overline{a}}}(1+
e^{\frac{\overline{b}- k\pi}{\overline{a}}})}{\overline{a}
\gamma_S},\label{mumax}
\end{equation}
where we have also multiplied by two in order to sum the (equal)
magnifications of the two images. In Sun retro-lensing, it amounts
to
\begin{equation}
\mu_{max}=1.29\times 10^{-14}
\left(\frac{M_L}{10M_\odot}\right)^2\left(\frac{R_S}{R_\odot}\right)^{-1}\left(\frac{D_{OL}}{0.01
pc }\right)^{-1}.
\end{equation}

Now, let us consider an arbitrary geometry, where observer, source
and lens are completely misaligned, so that $\gamma$ is far from
$k\pi$. Then we have to use the general formula (\ref{mu}) for the
magnification with a value for $\sin \gamma$ which is a generic
number between 0 and 1. To compare this situation with the perfect
alignment, we can take the ratio between Eq. (\ref{mu}) with $\sin
\gamma\simeq 1$ and one half of Eq. (\ref{mumax}) (for a single
image). Apart from the exponentials, which can weakly modify the
order of magnitude, this ratio is generally of the order of
\begin{equation}
\frac{\mu_{min}}{\mu_{max}}\sim\frac{\gamma_S}{2}\ll 1.
\end{equation}

Thus intermediate lensing is generally disfavored with respect to
standard and retro- gravitational lensing. However, this does not
mean that intermediate lensing cannot be interesting from an
observational point of view, as we shall see in Sect. 3.

\subsection{Time delay in strong field gravitational lensing}

Time delay in strong field gravitational lensing provides a very
important independent observable, which directly yields the
distance of the lens with a good accuracy \citep{BozMan}. In fact
the time delay between images with successive winding numbers is
\begin{equation}
\Delta T=2\pi \frac{D_{OL} \theta_m}{c_0}, \label{TDR}
\end{equation}
where $c_0$ is the speed of light. Thus, measuring $\theta_m$ and
$\Delta T$ we can immediately obtain $D_{OL}$. To get an idea of
how big the time delay is, we can write it as
\begin{equation}
\Delta T=0.13 h \left(\frac{M}{2.87 \times 10^6 M_\odot} \right).
\end{equation}

It is thus completely insignificant for solar mass black holes,
even if they are close to our solar system. It is also quite small
for the black hole at the center of our Galaxy, but it amounts to
several days for supermassive black holes in other galaxies, being
thus measurable, once we find a suitable variable source.

The calculations by \citet{BozMan} were made with the standard
gravitational lensing configuration in mind. However, it is easy
to see that in retro-lensing or any intermediate lensing
configurations the time difference between relativistic images
remains exactly the same, while only the (unobservable) absolute
travel time of the photon changes.

Now let us consider the time delay between the first relativistic
image and the direct image of the source. When the source is well
aligned behind the lens, then the direct image becomes weakly
lensed and the first relativistic image turns to the secondary
weak field image. In this case, the Shapiro time delay formula
applies (see \citet{WGS} for an application involving the Galactic
center).

If the source is not behind the lens, then the time delay between
the first relativistic image and the direct image is dominated by
the geometric path difference
\begin{eqnarray}
&\Delta T=& \frac{D_{OL}+D_{LS}-D_{OS}}{c}=\frac{D_{OL}}{c_0}
\left[ 1+\frac{\sin \beta -\sin \gamma}{\sin(\gamma-\beta)}
\right] \simeq
\nonumber \\
&& \simeq \frac{D_{OL}}{c_0}
\frac{1-\cos{\gamma}}{\sin{\gamma}}\beta, \label{TDSF-WF}
\end{eqnarray}
where $\beta$ is the angular separation between the source and the
black hole as seen by the observer. The last approximate equality
is valid if $\beta \ll 1$. In principle, measuring $\beta$ and
$\Delta T$, if we have a good knowledge of $\gamma$, we can use
the time delay to estimate the distance to the lens. With respect
to the time delay between consecutive relativistic images, this
measure is surely easier, since we just need one relativistic
image. However, we need a good knowledge of the angle $\gamma$,
which was not required in the previous case.

\section{Gravitational lensing of the star S2 by Sgr A*}

S2 is the star with the minimum average distance from the galactic
center discovered so far \citep{Sch1}. Its orbital motion has been
very accurately reconstructed through proper motion and spectral
measurements. Its orbital parameters, taken by \citet{Sch} are
reported in Table 1. This star looks like a O8-B0 main sequence
star of 15 M$_\odot$ with an apparent magnitude in the K-band
(centered on $\lambda=2.2 \mu$m) of $m_K=13.9$. The extinction in
the K band in the region of the galactic center amounts to 3.3
magnitudes \citep{Rie}. The orbital period and the major semiaxis
fix the enclosed mass to $M_{enc}=3.3\times 10^6 M_\odot$,
slightly larger than the central black hole mass, which is
currently estimated to $M_{BH}=2.87\times 10^6 M_\odot$.

\begin{table}
\centering
\begin{tabular}{cc}
\hline \hline
  Orbital parameter & Value \\
  \hline
  $a$ (pc) \dotfill & $4.54\times 10^{-3}$ \\
  $P$ (yr) \dotfill & 15.73 \\
  $e$  \dotfill& 0.87 \\
  $T_0$ (yr)  \dotfill& 2002.31 \\
  $i$ (deg)  \dotfill& 45.7 \\
  $\Omega$ (deg)  \dotfill& 45.9 \\
  $\omega$ (deg)  \dotfill& 244.7 \\
 \hline
\end{tabular}
\caption{Orbital parameters for S2. $a$ is the major semiaxis, $P$
is the orbital period, $e$ is the eccentricity, $T_0$ is the epoch
of periapse, $i$ is the inclination of the normal of the orbit
with respect to the line of sight, $\Omega$ is the position angle
of the ascending node, $\omega$ is the periapse anomaly with
respect to the ascending node. Data taken from \citet{Sch}.}
\end{table}

The value of the inclination of the orbit suggests that a high
alignment with the observer-lens line does not occur during the
motion of the star around Sgr A*. This seems to rule out any
possibility for standard or retro-gravitational lensing.
Contrarily to what expected, \citet{DepS2} claimed that the
relativistic images of S2 in the central black hole are not far
beyond instrumental sensitivities, even if the alignment is not
favorable. In this section, we complete their analysis in the
light of our formalism, including all the significant orbital
parameters and drawing light curves for the relativistic images
using our magnification formula. We can thus confirm their claim
also predicting the best observability time for these images.

Rather than a retro-lensing configuration, S2 represents a case
with an intermediate $\gamma$. The magnification of the images is
thus well described by Eq. (\ref{mu}). Since the radius of S2 is a
few solar radii, $\gamma_S$ stays much smaller than
$|\gamma-k\pi|$ (i.e. we are far enough from caustic points),
allowing us to trust the point-like magnification without need to
integrate it over the source surface. After some algebra, we can
write down all the interesting quantities in terms of the orbital
parameters of the system
\begin{eqnarray}
&& D_{LS}=\frac{a(1-e^2)}{1+e \cos \phi} \\
&& D_{OL}\simeq D_{OS}=8kpc\\
&& \gamma=\arccos[\sin(\phi+\omega)\sin i], \label{gammaS2}
\end{eqnarray}
where $\phi$ is the anomaly angle of the star starting from the
periapse epoch, $i$ is the inclination of the orbit and $\omega$
is the periapse anomaly with respect to the ascending node. By the
angular momentum conservation, we have
\begin{equation}
L= M_{S2}\sqrt{G M_{enc}a (1-e^2)}=M_{S2}D_{LS}^2 \dot \phi.
\end{equation}
By this equation, we can write a differential equation for $\dot
\phi$
\begin{equation}
\frac{\left[a(1-e^2) \right]^{3/2}}{\sqrt{G M_{enc}}(1+e \cos
\phi)^2}\dot \phi=1.
\end{equation}

\begin{figure}
\resizebox{\hsize}{!}{\includegraphics{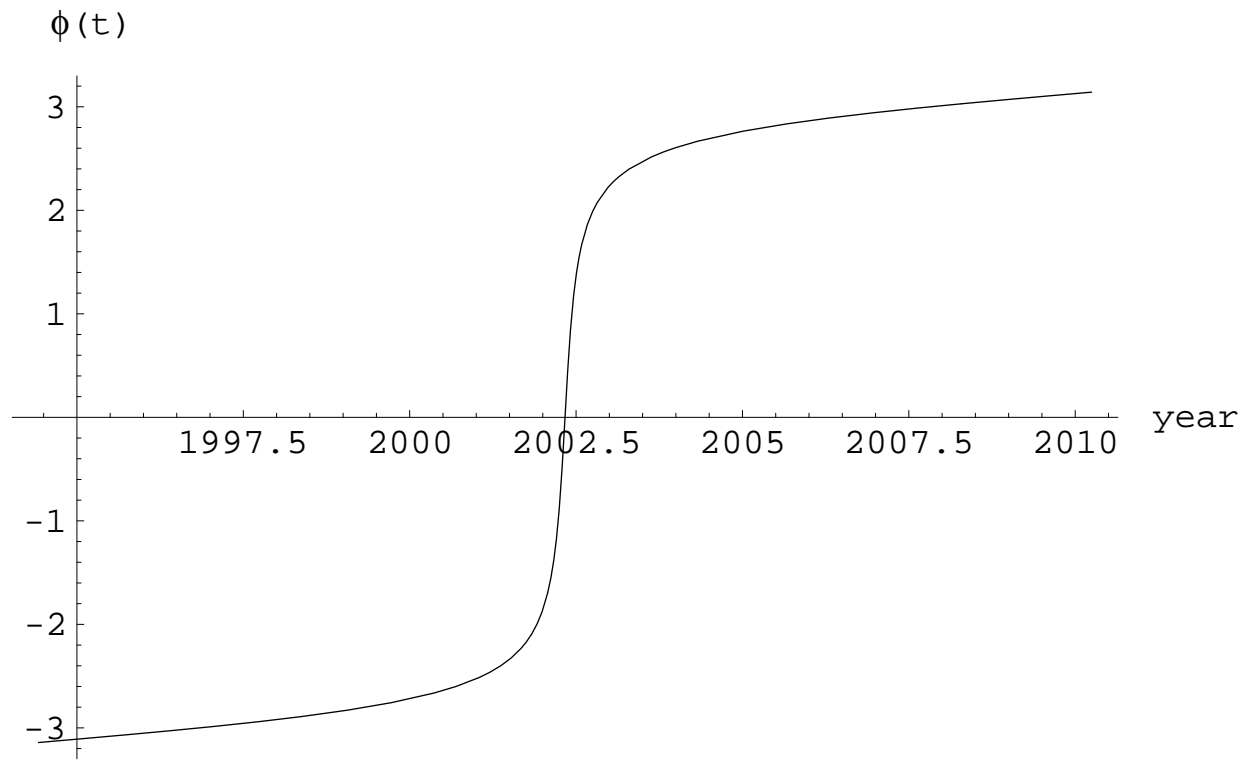}}
 \caption{The orbital position of S2 as a function of time.}
 \label{Fig phi}
\end{figure}

Integrating and inverting, we can get $\phi$ as a function of
time, exploiting the initial condition $\phi(T_0)=0$, with $T_0$
given in Table 1. If the eccentricity of the orbit of S2 were
negligible, $\phi(t)$ would just be a linear function of time, of
the form $\phi(t)=\omega_0 t$, with $\omega_0=2\pi/T$ being a
constant. This approximation was done for simplicity by
\citet{DepS2}. However, in Fig. \ref{Fig phi}, we see that the
high value of the eccentricity drastically modifies this function.
To get accurate predictions, it is thus mandatory to take into
account the angular motion of S2 correctly.

Finally, we can plot the magnification of the first two
relativistic images, taking $M_{BH}=2.87\times 10^6 M_\odot$ as
the mass of the black hole \citep{Sch} (\citet{DepS2} used
$M_{enc}$). The first relativistic image has $0<\gamma<\pi$. In
practice it is the secondary image of weak field gravitational
lensing which turns into a strong field image because of the high
misalignment. It has a negative parity and is formed close to the
black hole on the other side with respect to the direction of S2.
The second relativstic image has $\pi<\gamma<2\pi$ and comes from
light rays which take the "wrong" direction around the black hole,
and are bound to turn once more around it in order to reach the
observer. This image has positive parity and appears on the same
side of S2. The next images are fainter and probably uninteresting
for the moment.

\begin{figure}
\resizebox{\hsize}{!}{\includegraphics{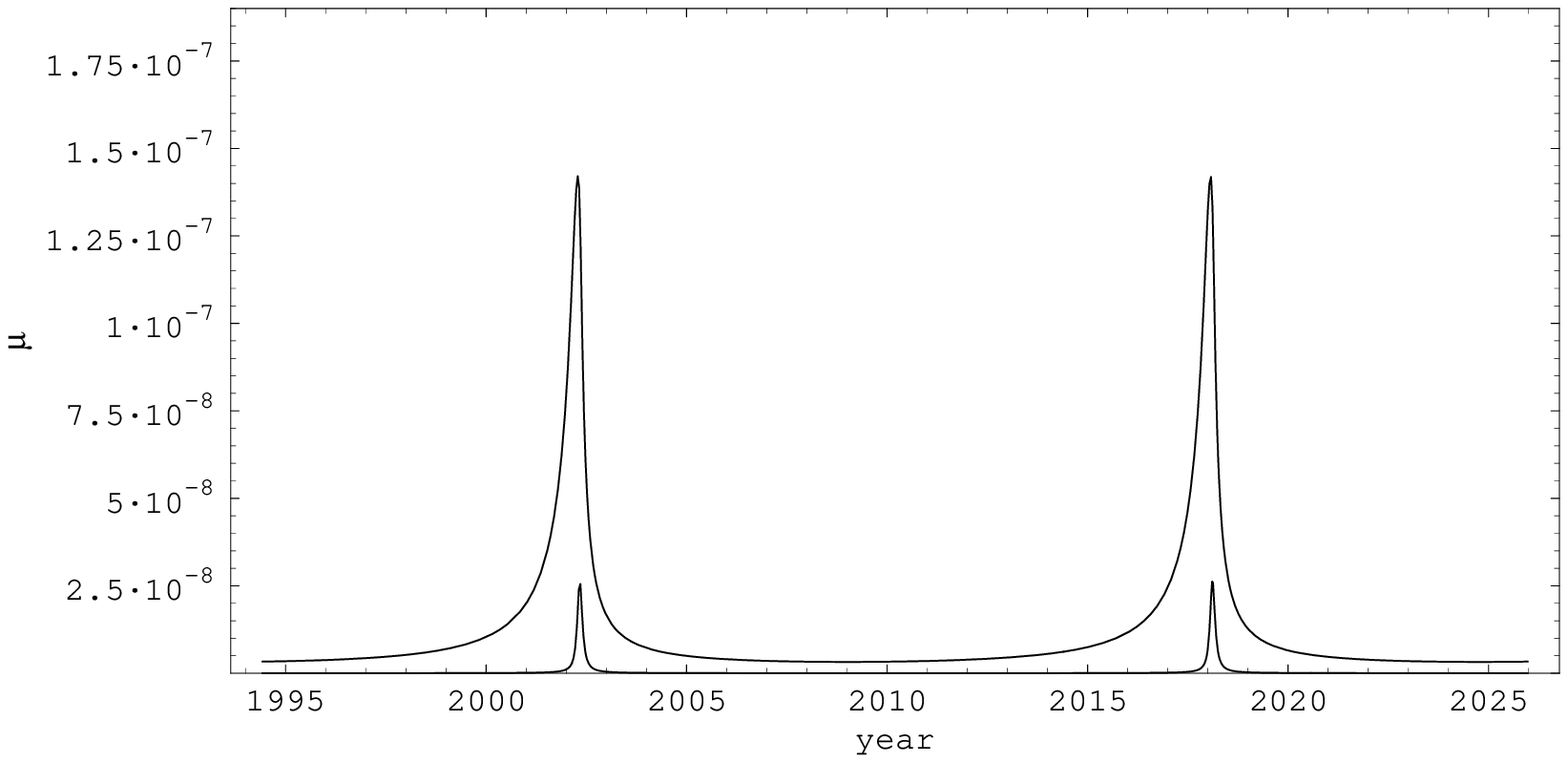}}
 \caption{Magnification of the first two relativistic images as a function of time.}
 \label{Fig S2mu}
\end{figure}

\begin{figure}
\resizebox{\hsize}{!}{\includegraphics{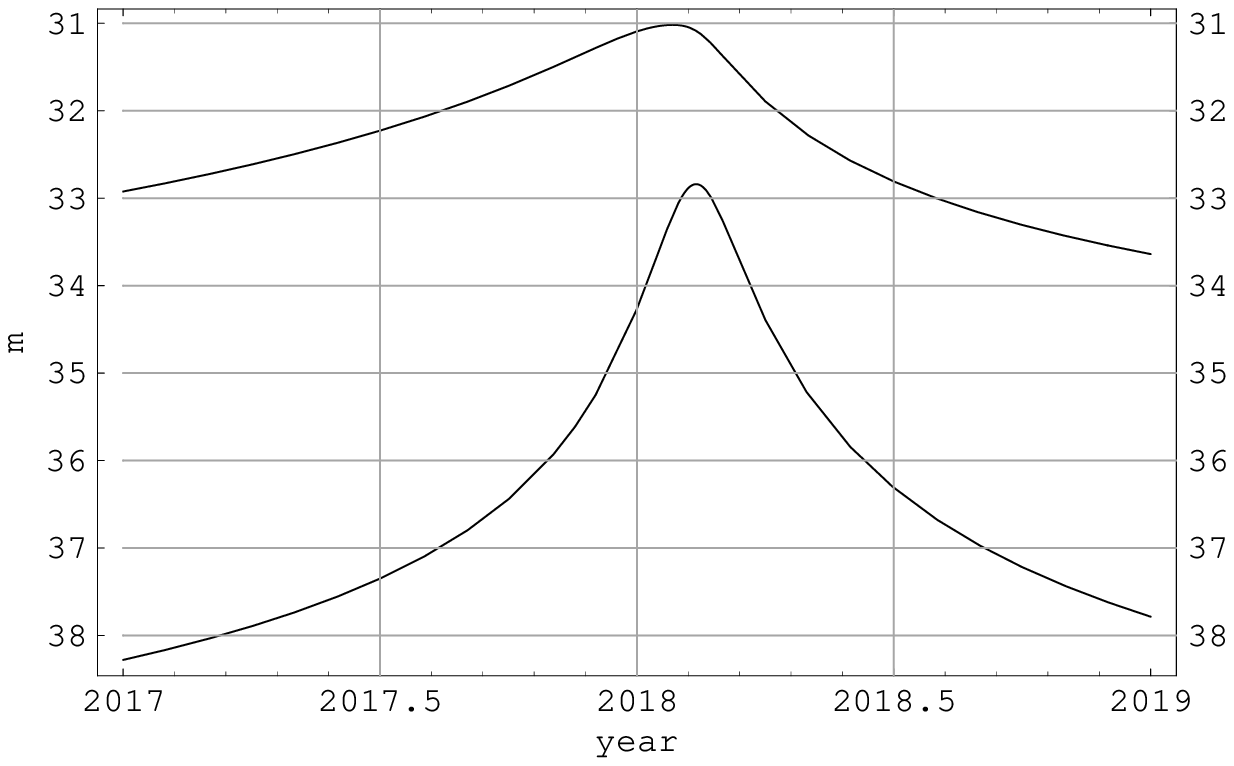}}
 \caption{Apparent magnitude of the first two relativistic images in the epoch of the next periapse.}
 \label{Fig S2ing}
\end{figure}

In Fig. \ref{Fig S2mu} we plot the magnifications of the first two
images as functions of time. The periapse epoch is the most
favorable for the observations since $D_{LS}$, which appears with
a power of -2 in Eq. (\ref{mu}), drops to its minimum value. This
minimum value is still 1300 times larger than the Schwarzschild
radius of the central black hole, so that we can safely treat the
source as far from the lens. Since the angular velocity of S2 is
maximal in the periapse epoch, the luminosity peak is relatively
short. As S2 passed through the periapse in year 2002, we have to
wait for the next periapse to get this luminosity peak. In Fig.
\ref{Fig S2ing}, we have drawn the expected apparent magnitudes in
the K band of the two first relativistic images at the epoch of
the next periapse, supposing that the extinction value 3.3 in this
band \citep{Rie} also applies from light coming to us from regions
very close to the central black hole. The first relativistic image
should stay brighter than 32 magnitudes from mid-August 2017 to
mid-April 2018. The second relativistic image is typically 5.4
times fainter during the peak. A hard challenge for next
generation instruments!

The observation of a relativistic image requires a strikingly high
angular resolution together with a high flux sensitivity. Very
impressive improvements in Very Long Baseline Interferometry have
been performed in the last years in the radio bands. With the
first detection of transatlantic fringes at 147 GHz \citep{Kri},
the present world record has been established at 18
microarceseconds, which is just few times the angular size of the
Schwarzschild radius of the black hole at the Galactic center.
Further improvements can be obtained at sub-mm wavelengths.

A very important step for absolute sensitivity in the infrared
band is the Next Generation Space Telescope (NGST), which will
operate in the wavelength interval $0.6 - 27 \mu m$. At
$\lambda=2.2 \mu m$, with a 3 hours exposure, it will be able to
detect a flux corresponding to 32 magnitudes, just enough to catch
the first S2 relativistic image. If coupled with other instruments
still to come, then the detection of relativistic images would
become a next future observational frontier. Furthermore, we
cannot forget that the Galactic center is surrounded by a crowded
clusters of stars, which can probably yield even better candidates
than S2 for strong field gravitational lensing.

We close this section with some comments on possible time delay
measurements. Since S2 is close to the center of Galaxy, but never
aligned behind it, we can estimate the time delay between the
first relativistic image and the direct image using Eq.
(\ref{TDSF-WF}). At the luminosity peak epoch, this delay amounts
to 5.95 hours, thus being comparable to the exposure time. In the
particular case of S2, no intrinsic variability has been reported
up to now in the observed spectral bands. However, the principle
of time delay measures could be applied to eventual new candidates
or to S2 itself if any variability is detected in its flux. A
measure of such a short time delay, requires a very good sampling
over the whole period of variability. The precision of such
estimate is limited by the effort undertaken to sample the two
light curves to be compared. With a good knowledge of the orbit
(in order to correctly estimate $\gamma$) and a sufficient
sampling, which would depend on the future facilities and the
characteristics of the source, it could be possible to establish
the distance to the center of the Galaxy on a solid direct
measurement.

\section{Kerr gravitational lensing}

What we have said up to now is true for spherically symmetric
black holes (including the estimates on S2 gravitational lensing).
In the case of spinning black holes, everything becomes much more
complicated, but some very important facts already emerge from the
analysis of quasi-equatorial motion, which has already been done
analytically \citep{Boz2}. In fact, it is evident that the
caustics are no longer aligned along the optical axis, but drift
following the sense of rotation of the black hole. In addition,
they are no longer point-like, but acquire finite extension. To
get an idea of the importance of these changes, we can see that
for $a=0.1$ the first retro-lensing caustic drifts from the
optical axis by an angle of $10^\circ$ and acquires an extension
of $2.4^\circ$ on the equatorial plane. The existence of extended
caustics is accompanied by the formation of pairs of additional
images when the source enters one of the caustics. These images
are missed in the quasi-equatorial approach since they rather live
far from the equatorial plane. A very important consequence is the
fact that high magnifications can be attained much more easily. In
fact, while in spherically symmetric metrics the caustics are
point-like and coincide with the optical axis, in the Kerr metric
they are distributed in all directions around the lens and have
finite extension, so that it is much easier for a source to lie
within a high amplification region for some relativistic images.
Another complication comes because the usual hierarchy among
relativistic images, which just follows the winding number, can be
completely upset by the fact that the source lies in a caustic
affecting images with higher winding number.

Summing up, the Kerr black hole lensing is much richer than
spherically symmetric black hole lensing. It is also much more
promising from a phenomenological point of view, since it is
easier to have brighter images. And of course it is reasonable to
expect that astrophysical black holes are born with a
non-negligible spin. All these statements point to the importance
of the investigation of the Kerr black hole lensing in the general
case.

\subsection{Kerr black hole lensing with S2}

As a practical example of the modifications that an eventual spin
of the black hole in Sgr A* would have on the light curve of the
images of S2, let us calculate the magnification at the moments
when S2 crosses the plane of the Galaxy. It is reasonable to
assume that if the central black hole is spinning, its equatorial
plane is very close to the rotation plane of the Galaxy. Using the
orbital parameters of S2, we see that it has crossed the galactic
plane in $t_1=2003.12$, where it had $\gamma=77.2^\circ$ and
$D_{LS}=2.74\times 10^{-3}pc$, being east of the black hole,
slightly behind it. It will cross again the galactic plane in
$t_2=2018.00$ with $\gamma=102.5^\circ$ and $D_{LS}=6.93\times
10^{-4}pc$, being west of the black hole, slightly before it. For
these crossing times, we can precisely calculate the
magnifications of the equatorial images using the formula by
\citet{Boz2}

\begin{eqnarray}
&\mu=&\frac{D_{OS}^2}{D_{OL}D_{LS}} \cdot\nonumber \\
&&\frac{\sqrt{u^2-a^2} u}{\overline{a} \left| \sqrt{u^2-a^2}
(D_{OL}+D_{LS}) C -D_{OL}D_{LS} S \right|}, \label{muKerr}
\end{eqnarray}
where $u=\theta R_{Sch}/D_{OL}$, $C=\cos \overline{\phi}$, $S=\sin
\overline{\phi}$ and $\overline{\phi}$ is the phase of the photon
oscillations on the equatorial plane, expressed by
\begin{equation}
\overline{\phi}=-\frac{\overline{b}-\gamma}{\overline{a}}+\hat{b}.
\end{equation}

The coefficients $\overline{a}$, $\overline{b}$ and $\hat b$ are
all functions of the spin of the black hole $a$, (see \citet{Boz2}
for the whole derivation).

\begin{figure*}
\resizebox{\hsize}{!}{\includegraphics{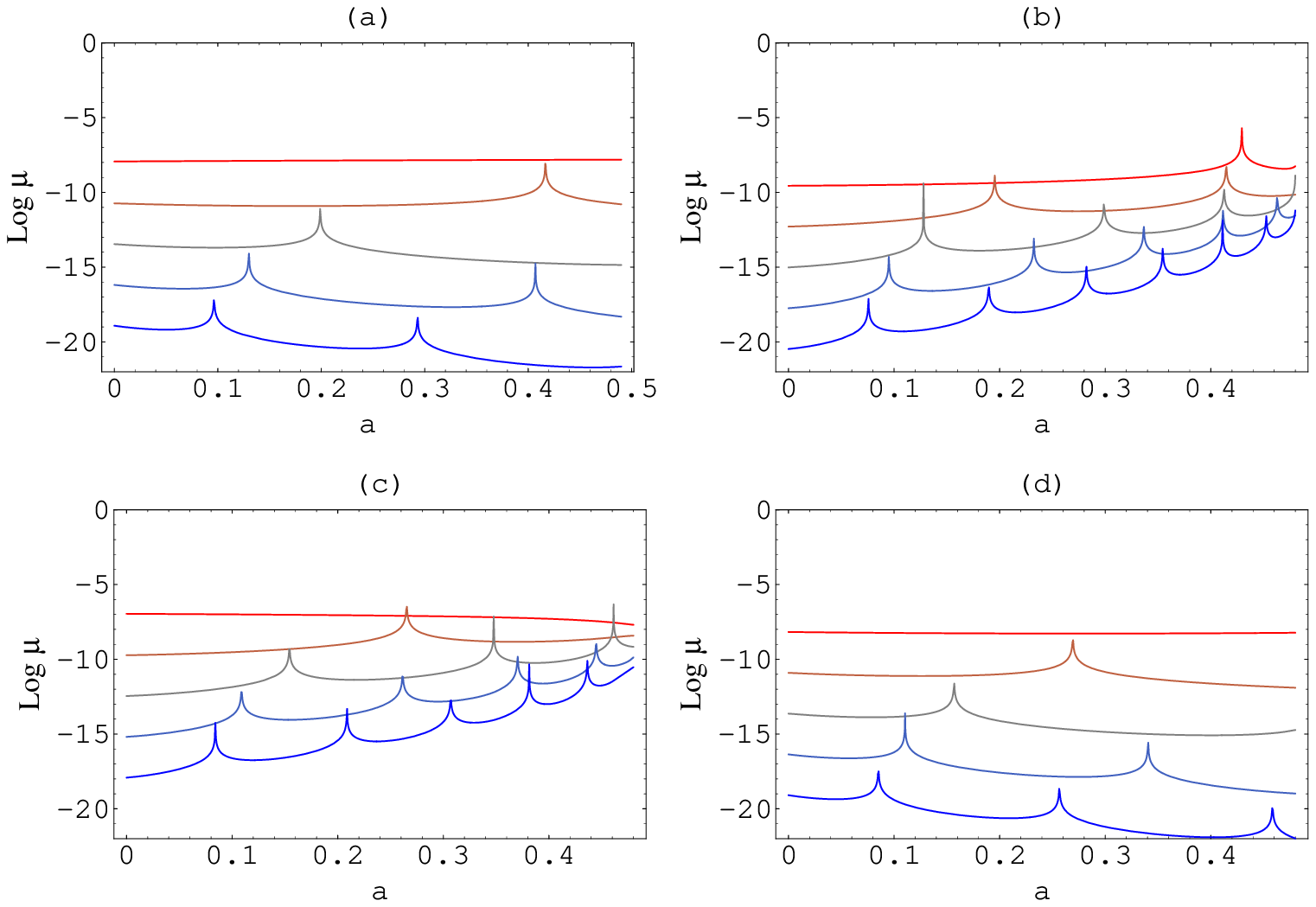}}
 \caption{Magnifications of the images of S2 as functions of the black hole spin $a$. (a) and (b) refer to the time
 $t_1=2003.12$, (c) and (d) refer to the time $t_2=2018.00$. (a) and (c) are the negative parity images with
 increasing winding numbers and (b) and (d) are the positive parity images, again with increasing winding number. }
 \label{Fig Kerr}
\end{figure*}

In Fig. \ref{Fig Kerr}, we plot the magnifications of the images
as a function of the black hole spin, from $a=0$ (Schwarzschild)
to $a=0.5$ (extremal Kerr black hole in our normalization). In
Figs. \ref{Fig Kerr}a and \ref{Fig Kerr}b, we plot the image
magnifications for the past time $t_1$, while in Fig. \ref{Fig
Kerr}c and \ref{Fig Kerr}d we plot the image magnifications for
the future time $t_2$. In the left figures we plot the
magnifications of the images with negative parity. The brightest
of them gives the most significant contribution (it is this image
that is going to be most likely observable). The magnifications of
the images with increasing winding number are plotted altogether.
The higher the winding number, the lower the magnification. On the
right figures we represent the magnification of the positive
parity images. The most relevant of these was also represented in
Figs. \ref{Fig S2mu} and \ref{Fig S2ing}, where it was however
fainter than the brightest negative parity image. Increasing the
winding number, the magnification falls down. The images in Figs.
\ref{Fig Kerr}a and \ref{Fig Kerr}d are formed by retrograde
photons, i.e. winding oppositely to the sense of rotation of the
black hole. The images in Figs. \ref{Fig Kerr}b and \ref{Fig
Kerr}c are formed by co-rotating photons.

Now let us comment on the outcome of these figures. Increasing the
spin of the black hole, we see that co-rotating images become
brighter and retrograde images become fainter. The most striking
features are the peaks in the magnification that occur at definite
values of the black hole spin. Varying the black hole spin, the
caustic points drift all around the trigonometric interval and for
some values they meet the angular position of S2. Around these
values, the magnification can become significantly higher than the
normal value. We also see that the first caustics, corresponding
to low winding numbers, move slowly and thus cross the position of
S2 less times. The caustics corresponding to higher winding
numbers move quickly. For this reason the fainter images meet
caustics more often in the interval of variability of $a$.

If the true value of the spin of the central black hole is such
that one of the images is close to a caustic point, that image
will be highly amplified. By these plots, we see that in year
2018.00, for the particular value $a\sim 0.26$ the second negative
parity image will overcome the brighter one.

Besides the luminosity of the relativistic images, the presence of
a spin for the black hole also affects their position. In fact,
image formed by retrograde photons appear farther from the central
black hole up to 1.5 times, while co-rotating photons appear
closer to it up to 2.6 times.

Finally, it must be kept in mind that since the caustics have
extended areas, these caustic points are actually just the cusps
of these caustics. This means that the images we are plotting in
Fig. \ref{Fig Kerr} change their parity at every caustic crossing.
Global parity conservation is assured by creation or destruction
of non equatorial images of the same parity of the image before
the crossing. For these images we do not have an analytical
treatment at the moment and all we can say is that they will
appear at each caustic crossing and disappear when the source
steps out of a caustic.

When a complete analytical treatment of the Kerr black hole
lensing is available, not only we will have a clearer idea of the
dynamics of all images, but we will be able to use the whole
information about position and luminosity of all visible
relativistic images to measure the spin of the black hole. We can
imagine that such an estimate would be much cleaner and simpler
than those relying on poorly understood models of the accretion
disk or gas surrounding the central black hole, which are subject
to very complicated physical processes and dynamics. However, for
the details of such a fascinating measure, we still have to wait
for further analytical progresses on the general Kerr lensing.

\subsection{Sun retro-lensing by a Kerr black hole}

Another context where Kerr black hole lensing was discussed was
retro-lensing of the Sun by a nearby Kerr black hole
\citep{DepRM}. As a first step, it was considered a geometry such
that the whole event occurred within the equatorial plane of the
black hole. Using the formulae by \citet{Boz2}, the positions of
the images were correctly calculated. However, the magnification
was estimated using the formula by \citet{HolWhe}, which was
derived for a Schwarzschild black hole (being also inaccurate for
central events as shown in Sect. 2). It is well known in different
contexts \citep{GouLoe,Bozp} that the modification to the
magnification map come firstly from the change to the Jacobian
function, due to the change in the lens model, and only secondly
from the displacements of the image positions due to the
modifications in the lens equation, which can be neglected in a
first extent. As a consequence of this simplification
\citet{DepRM} found no significant deviation from Schwarzschild
retro-lensing. However, by the study of the equatorial case, we
learn that the caustics drift and acquire finite extension, so
that we expect that things are not so simple.

For the configuration considered in \citet{HolWhe} and
\citet{DepRM}, we have strong magnification only if the source is
close to a caustic. Very small misalignments can already kill the
images to very small luminosities (it is sufficient to take a look
at Fig. \ref{Fig Micro} to get convinced; for quantitative
estimates see \citet{HolWhe}). So, it is sufficient that the
relevant caustic for retro-lensing moves out of the Earth orbit in
order that the possibility of Sun retro-lensing disappears at all.
With a black hole at distance 0.01 pc, this happens if the black
hole spin is $a>0.00027$: quite a small value. If we have a black
hole with a spin less than this, then the retro-lensing caustic
area would be much smaller then the Sun angular radius as seen
from the black hole, being effectively treatable as point-like.
Then everything would work more or less like in the Schwarzschild
case, with corrections to the magnification curves of the order of
$\delta\mu/\mu \simeq a$, i.e. below the precision of SFL
approximation.

\section{Conclusions}

Gravitational lensing in strong fields is a very interesting
subject from the theoretical point of view. Potentially it is a
very appealing phenomenon which is completely embedded within the
full general relativity. It would thus be an exceptional probe for
physics in the regions close to the event horizons of black holes
and would give very important feedback on the correct theory of
gravitation. This justifies the very strong efforts that have been
done by several groups to find possible candidate sources and
lenses which could make this fascinating phenomenon manifest. The
astrophysical cases investigated up to now are such that the
relativistic images should be "almost" observable or should become
observable within not many years. The most important problem is
whether these theoretical configurations are likely or not. The
case of S2, on the contrary, is a concrete case where definite
predictions for the image luminosities can be done.

In this work we have extended the analytical framework of the
Strong Field Limit analysis to cover all the possible geometric
configurations, giving up any small angle approximations which are
not automatically encoded in the tracks of the light rays. In this
way, we have been capable to revisit the so called retro-lensing,
in particular the Sun retro-lensing proposed by \citet{HolWhe},
with more valid mathematical instruments. Moreover, we can
adequately cover any intermediate case, such that of the star S2
\citep{DepS2}. For the first relativistic image of this star, we
predict an epoch of maximal luminosity (better than 32 magnitudes)
between the end of year 2017 and the first part of 2018.

Using previous work on equatorial Kerr lensing \citep{Boz2}, we
have guessed how the presence of extended caustics and their drift
from the optical axis can affect the brightness of the
relativistic images of S2 at the epoch of its crossing through the
galactic plane. We have also put upper limits on the black hole
spin in Kerr retro-lensing of the Sun in order to have still
significant images.

As a final remark, we can say that strong field gravitational
lensing opens really fascinating perspectives, like testing the
General Relativity, measuring the distance, the spin or other
physical parameters of the black hole in the Galactic center and
in the centers of other galaxies. The nice luminosity estimates
for S2 encourage to look for more stars orbiting very close to the
central black hole as new potential candidate sources for such an
amazing phenomenon.

\begin{acknowledgements}
We thank the CERN theory department for hospitality. We also thank
Gaetano Scarpetta for comments on the manuscript, Francesco De
Paolis and Achille Nucita for useful discussions, and our referee
for some nice suggestions.
\end{acknowledgements}

------------------------------------------------------------------

\bibliographystyle{aa}

\begin{thebibliography}{}
%


\bibitem[Atkinson(1965)]{Atk}  Atkinson, R.D. 1965, \aj, 70, 517

\bibitem[Bhadra(2003)]{Bha} Bhadra, A. 2003, \prd 67, 103009

\bibitem[Bozza(1999)]{Bozp}
Bozza, V. 1999, \aap, 348, 311

\bibitem[Bozza(2002)]{Boz1}
Bozza, V. 2002, \prd, 66, 103001

\bibitem[Bozza(2003)]{Boz2}
Bozza, V. 2003, \prd, 67, 103006

\bibitem[Bozza et al.(2001)]{BCIS}
Bozza, V., Capozziello, S., Iovane, G. \& Scarpetta, G. 2001, Gen.
Rel. and Grav., 33, 1535

\bibitem[Bozza \& Mancini(2004)]{BozMan}
Bozza, V., \& Mancini, L. 2004, Gen. Rel. and Grav., 36, 435

\bibitem[Chandrasekhar(1983)]{Cha}
Chandrasekhar, S. 1983, {\it Mathematical Theory of Black Holes},
Clarendon Press, Oxford

\bibitem[Cunningham \& Bardeen(1973)]{CunBar}  Cunningham, C.T., \& Bardeen, J.M. 1973, \apj, 183,
237

\bibitem[Dabrowski \& Schunck(2000)]{DabSch}  Dabrowski, M.P., \& Schunck, F.E. 2000, \apj, 535, 316

\bibitem[Darwin(1959)]{Dar}  Darwin, C. 1959, Proc. of the Royal Soc. of London, 249,
180

\bibitem[De Paolis et al.(2003)]{DepS2}
De Paolis, F., Geralico, A., Ingrosso, G. \& Nucita, A.A. 2003,
\aap, 409, 809

\bibitem[De Paolis et al.(2004)]{DepRM}
De Paolis, F., Geralico, A., Ingrosso, G., Nucita, A.A. \& Qadir,
A. 2004, \aap, 415, 1

\bibitem[Eiroa, Romero \& Torres(2002)]{ERT} Eiroa, E.F., Romero, G.E., \& Torres, D.F. 2002, \prd, 66,
024010

\bibitem[Eiroa \& Torres(2003)]{EirTor}
Eiroa, E.F. \& Torres, D.F. 2003, gr-qc/0311013

\bibitem[Frittelli, Kling \& Newman(2000)]{FKN} Frittelli, S., Kling, T.P., \& Newman, E.T. 2000, \prd, 61, 064021

\bibitem[Frittelli \& Newman(1999)]{FriNew} Frittelli, S., \& Newman, E.T. 1999, \prd, 59, 124001
%

\bibitem[Gould \& Loeb(2002)]{GouLoe} Gould, A., \& Loeb, A.
1992, \apj, 396, 104

\bibitem[Holz \& Wheeler(2002)]{HolWhe}
Holz, D.E. \& Wheeler, J.A. 2002, \apj, 578, 330

\bibitem[Krichbaum et al.(2002)]{Kri}  Krichbaum, T.P., Graham, D.A., Alef, W., et al. 2002,
Proc. of the 6th European VLBI Network Symposium

\bibitem[Luminet(1979)]{Lum}
Luminet, J.P. 1979, \aap, 75, 228

\bibitem[Nemiroff(1993)]{Nem}
Nemiroff, R.J. 1993, Amer. Jour. Phys., 61, 619

\bibitem[Ohanian(1987)]{Oha}
Ohanian 1987, Amer. Jour. Phys., 55, 428

\bibitem[Perlick(2003)]{Per}  Perlick, V. 2003, gr-qc/0307072

\bibitem[Petters(2003)]{Pet} Petters, A.O. 2003, \mnras, 338, 457

\bibitem[Rieke, Rieke \& Paul(1989)]{Rie}  Rieke, G.H., Rieke, M.J., \& Paul, A.E. 1989, \apj, 336,
752

\bibitem[Sch\"odel et al.(2002)]{Sch1} Sch\"odel, R., Ott, T., Genzel, R., et al.  2002, \nat, 419,
694

\bibitem[Sch\"odel et al.(2003)]{Sch} Sch\"odel, R., Ott, T., Genzel, R., et al.  2003, \apj, 596,
1015

\bibitem[Vazquez \& Esteban(2003)]{VazEst} Vazquez, S.E., \& Esteban, E.P. 2003, gr-qc/0308023.

\bibitem[Viergutz(1993)]{Vie}
Viergutz, S.U. 1993, \aap, 272, 355

\bibitem[Virbhadra \& Ellis(2000)]{VirEll}
Virbhadra, K.S., \& Ellis, G.F.R. 2000, \prd, 62, 084003

\bibitem[Wex, Gil \& Sendyk(1996)]{WGS}
Wex, N., Gil, J. \& Sendyk, M. 1996, \aap, 311, 746

\bibitem[Witt \& Mao(1994)]{WitMao}
Witt, H.J., \& Mao, S. 1994, \apj, 430, 505

\end{thebibliography}

\end{document}